\documentclass[10pt,aps,amsmath,amssymb,reprint,citeautoscript,floatfix,superscriptaddress]{revtex4-2}
\usepackage[usenames,dvipsnames]{color}
\usepackage{graphicx,xfrac,microtype}
\usepackage[bookmarks=false,colorlinks]{hyperref}
\hypersetup{hypertexnames=false,linkcolor=RubineRed,citecolor=ForestGreen,filecolor=Mulberry,urlcolor=RoyalBlue}
\usepackage{algorithm} 
\usepackage{algpseudocode}

\begin{document}

\title{Contour Exploration with Potentiostatic Kinematics}

\author{Michael J.\ Waters}
\email{michael.j.waters@northwestern.edu}
\affiliation{Department of Materials Science and Engineering, 
Northwestern University, Evanston, Illinois 60208, USA}

\author{James M.\ Rondinelli}\email{jrondinelli@northwestern.edu}
\affiliation{Department of Materials Science and Engineering, 
Northwestern University, Evanston, Illinois 60208, USA}

\date{\today}

\begin{abstract}
We introduce a method of exploring potential energy contours in complex dynamical systems based on potentiostatic kinematics wherein the systems are evolved with minimal changes to their potential energy.  
We construct a simple iterative algorithm for performing potentiostatic kinematics, which uses an estimate curvature to predict new configurations space coordinates on the potential energy contour and a potentiostat term component to correct for errors in prediction.
Our methods are then applied to atomic structure models using an interatomic potential for energy and force evaluations as would commonly be invoked in a 
molecular dynamics simulation. 
Using several model systems, we assess the stability and accuracy of the method on  different hyperparameters in the implementation of the potentiostatic kinematics.
Our implementation is open source and available within the Atomic Simulation Environment (ASE) package for atomic simulation. 
\end{abstract}

\maketitle

\section{Introduction}
When constructing surrogate models of complex dynamical systems, it is convenient to use sets of time evolution sequences to sample realistic parameter spaces. 
For some complex models, sequential sampling yields reduced computational cost due to inherent conditioning of iterative solvers. 
For example, the self-consistency cycle of a density functional theory calculation is greatly accelerated by reusing the electron density and wave functions from previous structural iterations.
Likewise, dynamical simulations can often be tuned to be well behaved so as not to produce numerically unstable or unphysical results that must be avoided when attempting to explore parameter spaces \textit{a priori}. 
Molecular dynamics (MD) fulfills the desirable traits of being sequential and does not produce unstable or unphysical results when performed with a reasonable time step.
Recently MD simulations have begun making extensive use of  machine learned interatomic potentials (MLIPs) constructed from \textit{ab initio} data \cite{MuellerMLIP}.
Sampling data for the MLIP sourced from realistic dynamics would seem to yield reasonable training data, but there are several issues with this approach: 
(1) Rare events are, by definition, undersampled. 
(2) Time integration requires a choice of time steps. If the steps are too small then the data is not unique; if they are too large then the results can become unphysical.
(3) A lack of inherent applicability limits. In most dynamical simulations, kinetic energy, potential energy, temperature, pressure, \textit{etc.} vary so there is no easy way describe the limits of the training set. 

To address these issues, we propose contour exploration based potentiostatic kinematics as a complementary method to obtain sampling data for MLIPs. 
In the contour exploration scheme, dynamical systems are evolved following their potential energy contours. 
Conceptually, it is equivalent to circumnavigating a mountain by traversing only a particular elevation. 
Traditional dynamics would have a rock rolling downhill.
When encountering a ravine, the rock my roll uphill for a short while before following a new downhill direction.
Energy minimization methods would have the rock jump to the bottom in as few of steps as possible.
Here we use potentiostatic kinematics to explore the potential energy surface (PES) by following the potential energy contours (PECs).
By doing so, we sample configuration space in a manner distinct from dynamical simulations, which may improve sampling of rare events or, at least serves to augment traditional simulations. 
By controlling the potential energy, training data can be generated within a specific potential energy range. 
Any models trained on or fit to this data will carry that energy range as clear limits of their applicability, \textit{i.e.}, it's easy to determine when the model is being used outside its energy window and this window is easily determined and communicated.  
Finally, there is no need for a time step in the traditional sense, because motion occurs only perpendicular to the potential energy gradients. 
Instead, the step size is determined by the PEC radius of curvature. Since no step can perfectly land on the PEC target energy, we introduce a potential energy analogue of a thermostat, a potentiostat, which acts to maintain a target potential energy. 
In the following work, we describe and demonstrate a minimal algorithm  
for potentiostatic kinematics, which when paired with an on-contour momentum drift term, can be used for contour exploration.
The application of this technique for   
creating training data sets for \emph{ab initio} MLIPs will be explored in the future.  

\section{Theory}
With the  goal of creating large position updates with minimal changes in potential energy, the total velocity vector, $\mathbf{V}$, must be maintained perpendicular to the  total force vector, $\mathbf{F}$, such that 
$\mathbf{F} \cdot \mathbf{V} = 0$.
%
By performing a vector rejection of the velocities on the forces and re-scaling on the resulting velocity vector, the kinetic energy can be conserved. 
For a non-stationary reference frame or a system with net velocity, the net system velocity can be conserved by excluding it from the vector rejection and adding it to the resulting vector.
However, unless very small position updates, $\Delta \mathbf{r}$, are performed, the potential energy will quickly wander from the target. 
Therefore, we introduce a potentiostat term, $\Delta\mathbf {r}_{\perp }$ to our position update, $\Delta\mathbf{r}$, that is perpendicular to our velocity vector and along the force vector to correct for deviations from the target energy as 
$\Delta\mathbf{r} =  \Delta\mathbf{r}_{\parallel } + \Delta\mathbf{r}_{\perp }$, 
where $\Delta\mathbf{r}_{\parallel }$ is the position update for the contour-following step component given by the velocity vector along the contour.
The potentiostat term can be computed by linear approximation:
\begin{equation}
 \Delta\mathbf{r}_{\perp } =  \frac{U-U_{target}}{\left | \mathbf{F} \right |} \frac{\mathbf{F}}{\left | \mathbf{F} \right |}\,.
\end{equation}
With these two components, a stable potentiostatic contour exploration algorithm can be constructed. Although a better approach is to use the curvature of the potential energy contour to make better predictions for the position update. For this purpose, we employ Frenet–Serret formulas \cite{DiffGeom}:
\[ \begin{aligned}
\dfrac{d\mathbf{T}}{ds} &= \kappa\mathbf{N}, \\
\dfrac{d\mathbf{N}}{ds} &= -\kappa\mathbf{T}+\tau\mathbf{B},\\
\dfrac{d\mathbf{B}}{ds} &= -\tau\mathbf{N},
\end{aligned} \]
where $\mathbf{T}$ is the unit tangent vector and is also the normalized velocity vector along the direction of motion, $s$ is the trajectory arc length, $\kappa$ is the curvature,  $\mathbf{N}$ is the normal unit vector and is also the force unit vector, $\tau$ is the torsion, and $\mathbf{B}$ is the binormal unit vector given as the cross product of $\mathbf{T}$ and $\mathbf{N}$. 
The curvature can be computed as:
\begin{equation}
\left \|  \frac{ d \boldsymbol{\mathbf{N}} }{d s}  \right \|  = \kappa    
\end{equation}
The Taylor expansion of the Frenet–Serret equations \cite{DiffGeom} yields a practical formula for the contour following position update:
\[\begin{aligned} &\Delta\mathbf {r}_{\parallel } (s)  =\\  &\left(s-\frac{s^3\kappa^2}{6}\right)\!\mathbf T  + \left(\frac{s^2\kappa}{2}+\frac{s^3\kappa'}{6}\right)\!\mathbf N + \left(\frac{s^3\kappa\tau}{6}\right)\!\mathbf B + O(s^4)\,.   
\end{aligned}\]
By assuming locally constant curvature, the Taylor expansion can be approximated as:
\begin{equation}
\label{equ_fs_update}
  \Delta\mathbf {r}_{\parallel } (s) \approx   \left(s-\frac{s^3\kappa^2}{6}\right)\mathbf T + \left(\frac{s^2\kappa}{2}\right)\mathbf N   
\end{equation}
so that only the previous position and forces are needed to calculate the curvature. 
Likewise, the potentiostat term can be improved by extrapolating at constant curvature:
\begin{equation} \label{equ_normal_extrapolation}
\mathbf{N}_{new} = \mathbf{N}_{current} + \frac{d \boldsymbol{\mathbf{N}} }{d s} s    
\end{equation}
\begin{equation} \label{equ_potentiostat}
\Delta\mathbf{r}_{\perp } =  \frac{U-U_{target}}{\left | \mathbf{F} \right |} \mathbf{N}_{new}   
\end{equation}
Using the curvature, the step size can be automatically tuned by the contour direction change angle, $\phi$ using the chord length:
\begin{equation}
\label{equ_chord}
| \Delta \mathbf{r} | = \kappa^{-1} \sqrt{2-2 \cos \phi }\,,
\end{equation}
which is useful since step size can be controlled by a scale-free parameter.

For data generation in MLIP construction, it is often desirable to explore the largest space possible to ensure the interatomic potential is robust to describe different bonding configurations.
Within purely potentiostatic kinematics, it is possible for a system to enter cyclic orbits. 
One simple example of this occurs for the case of two atoms orbiting each other. Another example is where a highly symmetric structure undergoes dilation cycles with velocity vectors of the same symmetry. %
In either case, a drift term, $\Delta \mathbf{r}_{drift}$, can be added to position update as  
\begin{equation} \label{equ_update_components}
\Delta\mathbf {r} = \Delta \mathbf{r}_{\perp} + \Delta \mathbf{r}_{\parallel }\left ( s \right ) + \Delta \mathbf{r}_{drift}
\end{equation}
to tune these trajectories. 
To eliminate redundancy with the contour-following term, $\Delta \mathbf{r}_{\parallel}$ and to conserve energy, 
the drift term should be orthogonal to both the contour-following term and the potentiostat term, $\Delta \mathbf{r}_{\perp}$, as follows: 
\[ \mathbf{N}_{new} \cdot \Delta \mathbf{r}_{drift} = 0 \]
\[ \mathbf{T}_{new} \cdot \Delta \mathbf{r}_{drift} = 0 \]
\section{Practical Implementation}
The aforementioned contour exploration scheme is implemented as a \texttt{Dynamics} module for the Atomic Simulation Environment (ASE) package.\cite{ASE} 
Since its standard calculator interface permits the modular usage with many popular MD and \textit{ab initio} codes without modification, the contour-exploration module can be used in conjunction with any of the ASE motion constraints. 
Efforts are underway to include the module within ASE.
The current implementation enforces a maximum step size. 
%
We use step projections, $s_{\parallel }$, $s_{\perp}$, $s_{drift}$,  along the three orthogonal step components (\autoref{equ_update_components}) so that the net step size can be maintained regardless of rotation in the unit tangent, $\mathbf{T}$, and the unit normal direction, $\mathbf{N}$:
%
\[\begin{aligned}
\Delta \mathbf{r}_{\parallel } &= s_{\parallel } \,  \mathbf{T}_{new}\,,\\ 
\Delta \mathbf{r}_{\perp }     &= s_{\perp}      \,   \mathbf{N}_{new}\,, \\
\Delta \mathbf{r}_{drift }     &= s_{drift}      \,  \mathbf{D}\,,     \\ 
s^{2} &= \left ( \Delta \mathbf{r} \right )^{2} = s_{\parallel }^{2} + s_{\perp }^{2} + s_{drift}^{2}\,, 
\end{aligned}
\]
%
where $s$ is the total desired step size, and $\mathbf{D}$ is the drift unit vector. 
Ideally, potentiostat and drift terms are small adjustments to the contour-following term so that they function more as perturbations rather than dominating the dynamics. 
This scenario may not always be possible when far from the target energy, so priority is given to potentiostatic component of the position update.

\begin{figure}[t]
\begin{algorithm}[H]
	\caption{\label{fig:algo}Contour Exploration} 
	\begin{algorithmic}
		\For {$stepnum=1,2, \ldots Nsteps$}
		    \State Update normal and tangent direction from current forces
		    
            $ \mathbf{N}  \leftarrow  \mathbf{F}/  | \mathbf{F} | $
            
            $ \mathbf{T}  \leftarrow (\mathbf{V} \, \mathrm{reject} \, \mathbf{N})/| \mathbf{V} \, \mathrm{reject} \, \mathbf{N} | $
            
            \State  Use the previous position update to calculate normal derivative, tangent derivative, and curvature
            
            $ \dfrac{d\mathbf{N}}{ds} \leftarrow \Delta \mathbf{N} / | \Delta \mathbf{r}_{old} | $
                        
            $ \dfrac{d\mathbf{T}}{ds} \leftarrow \Delta \mathbf{T} / | \Delta \mathbf{r}_{old} | $

            $\kappa \leftarrow    \left |   \dfrac{d\mathbf{N}}{ds} \right |    $
            
            \State Calculate the step size as the smaller of the chord of the angle limit or the maximum step size
            
            $s \leftarrow  \min \left( \frac{ \sqrt{2-2 \cos \phi } }{ \kappa }, s_{max} \right ) $
            
            \State Compute the potentiostat step size
            
            $s_{\perp} \leftarrow  \alpha  \frac{U-U_{target}}{\left | \mathbf{F} \right |} $
            
            \State Compute the contour-following and drift-step sizes  

            $ s^{2}_{remain} \leftarrow  s^{s} - s_{\perp }^{2} $
            
            $ s_{\parallel }^{2} \leftarrow \left ( 1 -\beta^{2}  \right)    s^{2}_{remain} $
            
            $ s_{drift }^{2} \leftarrow \beta^{2}    s^{2}_{remain} $

            \State Estimate the new normal and tangent vectors
            
            $ \mathbf{N}_{new} \leftarrow \mathbf{N} + \frac{\mathrm{d}\boldsymbol{\mathbf{N}} }{\mathrm{d} s} s_{\parallel} $
            
            $ \mathbf{T}_{new} \leftarrow \mathbf{T} + \frac{\mathrm{d}\boldsymbol{\mathbf{T}} }{\mathrm{d} s} s_{\parallel} $
            
            $ \mathbf{N}_{new} \leftarrow \mathbf{N}_{new} / \left |  \mathbf{N}_{new}  \right| $
            
            $ \mathbf{T}_{new} \leftarrow \mathbf{T}_{new} / \left |  \mathbf{T}_{new}  \right| $
            
            \State Compute the contour-following step using the truncated Taylor expansion
            
            $  \Delta\mathbf {r}_{\parallel } (s) \leftarrow   \left(s-\frac{s^3\kappa^2}{6}\right)\mathbf T + \left(\frac{s^2\kappa}{2}\right)\mathbf N   $
            
            \State Compute the potentiostat correction using the estimated normal direction
            
            $  \Delta \mathbf{r}_{\perp }     = s_{\perp}      \,   \mathbf{N}_{new} $
            
            \State Create random drift vector perpendicular to estimated normal and tangent direction
            
            $\mathbf{D}  \leftarrow    \mathrm{RandomVector} ( N_{atoms}, 3 )  $
            
            $ \mathbf{D}  \leftarrow    \mathbf{D} \, \mathrm{reject} \, \mathbf{N_{new}} $
            
            $ \mathbf{D}  \leftarrow    \mathbf{D} \, \mathrm{reject} \, \mathbf{T_{new}} $

            $  \mathbf{D}  \leftarrow  \mathbf{D}/  | \mathbf{D} | $ 
            
            $   \Delta \mathbf{r}_{drift }   \leftarrow  s_{drift}      \,  \mathbf{D}       $
            
			\State Sum contributions and update positions respecting constraints
			
			$\Delta\mathbf {r} = \Delta \mathbf{r}_{\perp} + \Delta \mathbf{r}_{\parallel }\left ( s \right ) + \Delta \mathbf{r}_{drift} $
			
			$ \mathrm{UpdatePositions}(\Delta\mathbf {r}) $
			
		\EndFor
	\end{algorithmic} 
\end{algorithm}
\end{figure}
The position update cycle starts with computing the normal vector, $\mathbf N$, from the forces as shown in \textcolor{RubineRed}{Algorithm}~\autoref{fig:algo}.
The tangent vector, $\mathbf T$, is updated from the normalized velocity vector rejection of the normal vector. 
The arc length derivatives of the normal vector and the tangent vector are estimated using the magnitude previous position update as the change in the arc length. 
The curvature is calculated as the norm of the arc length derivative of the normal vector since it is directly from the PES and while the tangent vector will inherit noise if drift is used. 
The step size is determined using the chord length formula (\autoref{equ_chord}) and the angle limit $\phi$, which is restricted to a maximum step, $s_{max}$.
Next, the potentiostat step size, $s_{\perp}$, is calculated by \autoref{equ_potentiostat} but in practice we apply a scaling factor, $\alpha$, that determines how aggressively deviations from the target energy are corrected by motion parallel to the forces. 
When $\alpha < 1$, the potentiostat term shifts the system only partially  
towards the extrapolated target energy location. 
When $\alpha > 1$, the potentiostat term overshoots the extrapolated target energy location. 
The potentiostat step size can be either positive or negative depending on whether the system needs to move with or against the  potential energy gradient in order to reach the energy target.
Next, the potentiostat step size is subtracted step size budget. 
The drift step size is fixed to a fraction, $\beta$ of the remaining step size budget and the rest is applied to the contour following step size. 
The new normal vector and new tangent vectors are estimated with the linear extrapolation of formula in \autoref{equ_normal_extrapolation} using the contour following step size and then renormalized.

The new normal vector is multiplied by the potentiostatic step projection to give the potentiostatic component of the position update. 
The contour-following component of the position update is calculated with \autoref{equ_fs_update} using the contour-following step size as the arc length.
To create the drift component of the position update, a random vector is made perpendicular to the new normal vector and the new tangent vector and then scaled in magnitude to the drift step size. 
The net position update is finally a sum of: the contour following term, potentiostatic term, and drift term.  
Once the position update is passed through/filtered by a constraint system, such as RATTLE \cite{RATTLE}, the new velocity vector is assigned by the permitted position update.

\begin{figure}
\centering
\includegraphics[width=0.3\textwidth,clip]{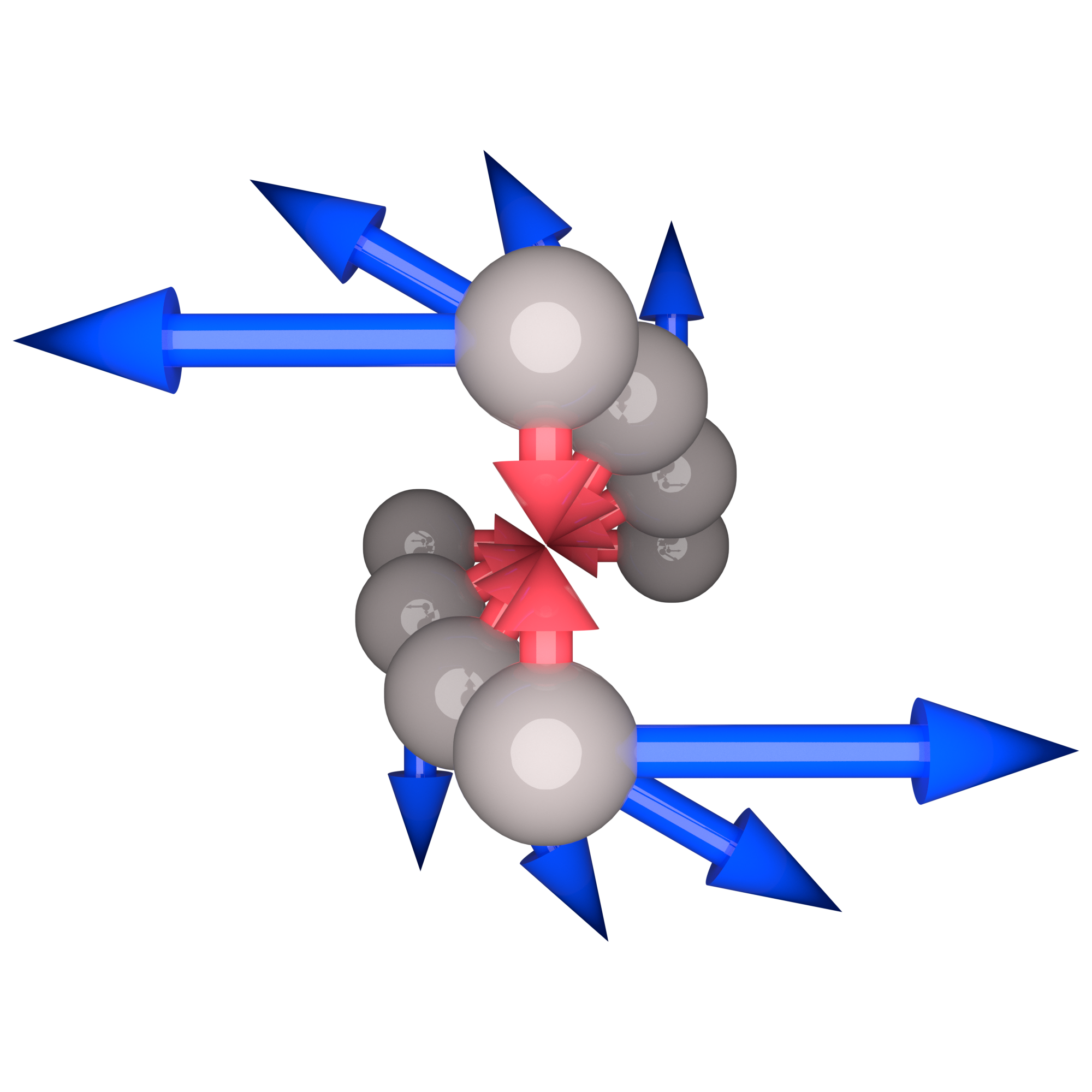}
\caption{\label{fig:two_atoms} The initial conditions, and the first three iterations are stacked, coming out of the page. 
The velocity vectors are blue and the force vectors are red. 
The 30$^\circ$ rotations confirm that step sizes are scaled to angle as intended. 
}
\end{figure}

\begin{figure*}
\centering
\includegraphics[width=0.3\textwidth,clip]{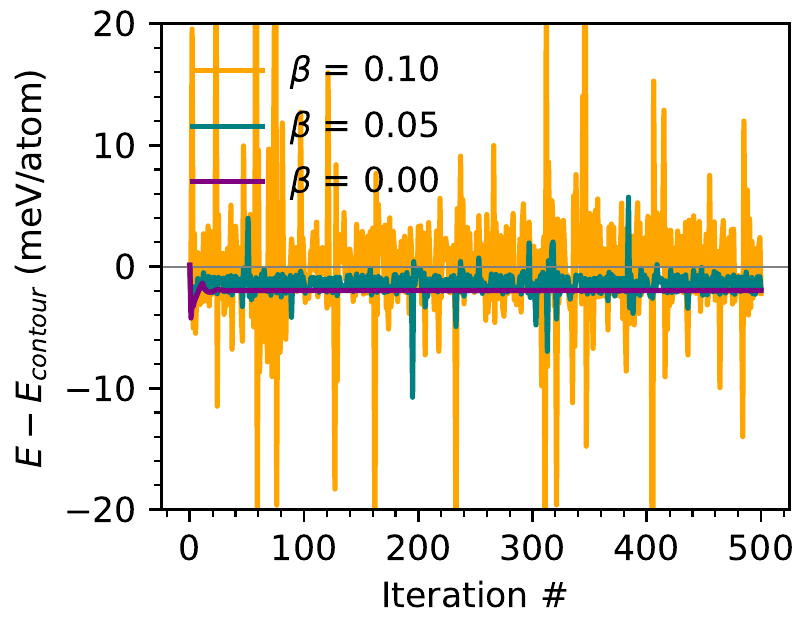}
\includegraphics[width=0.3\textwidth,clip]{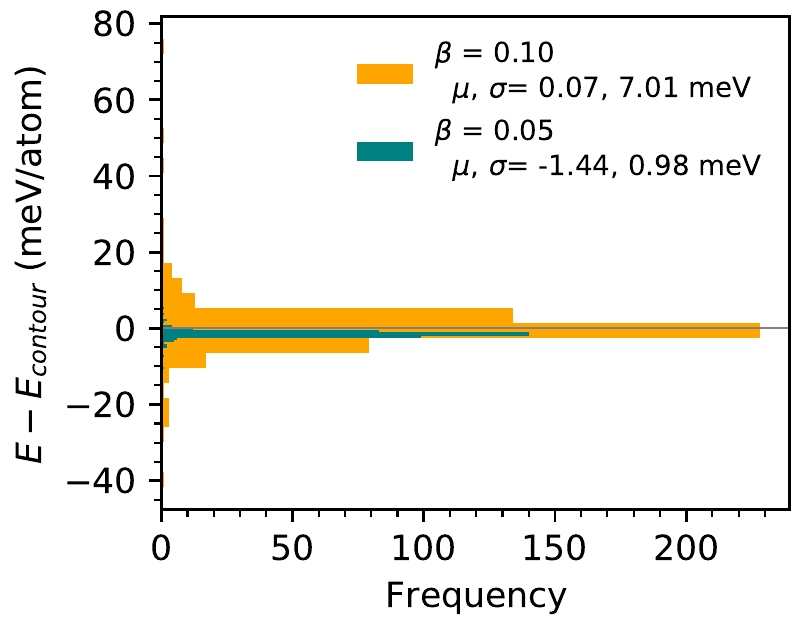}
\includegraphics[width=0.3\textwidth,clip]{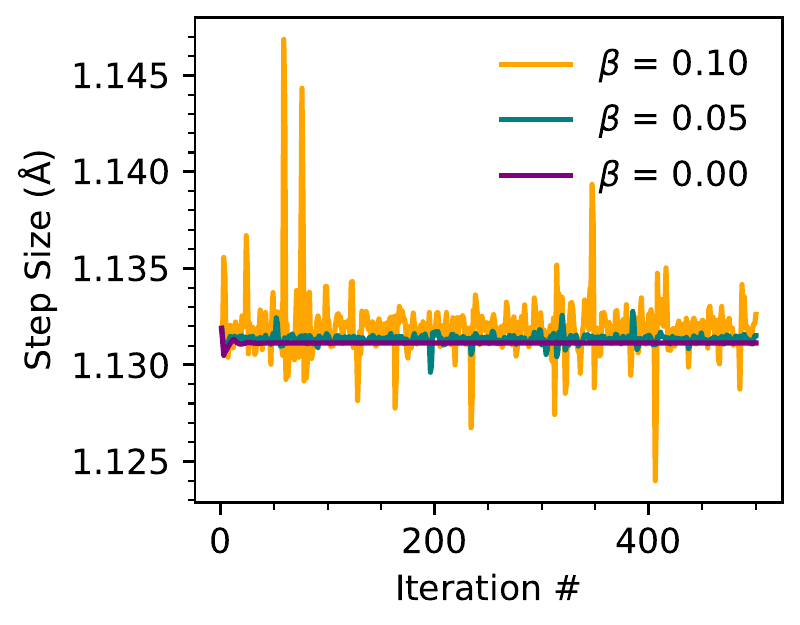}
\includegraphics[width=0.3\textwidth,clip]{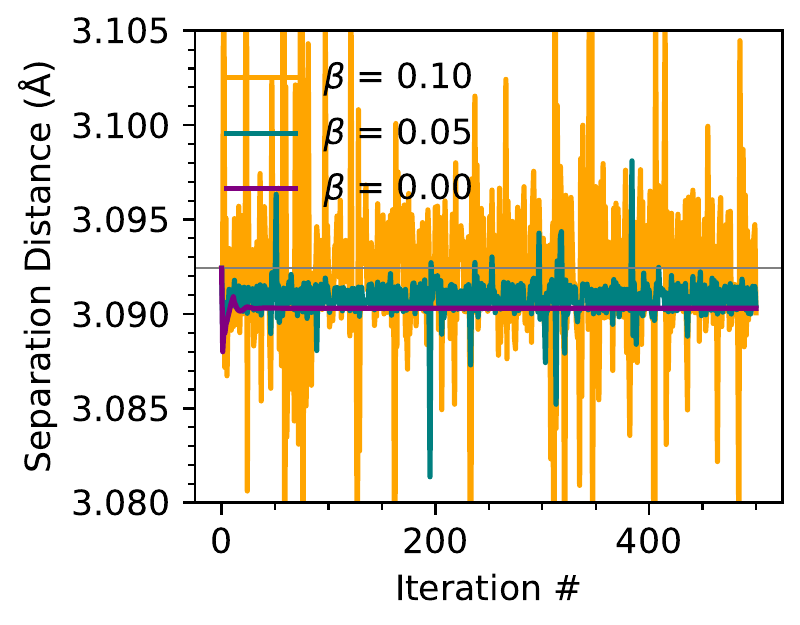}
\includegraphics[width=0.3\textwidth,clip]{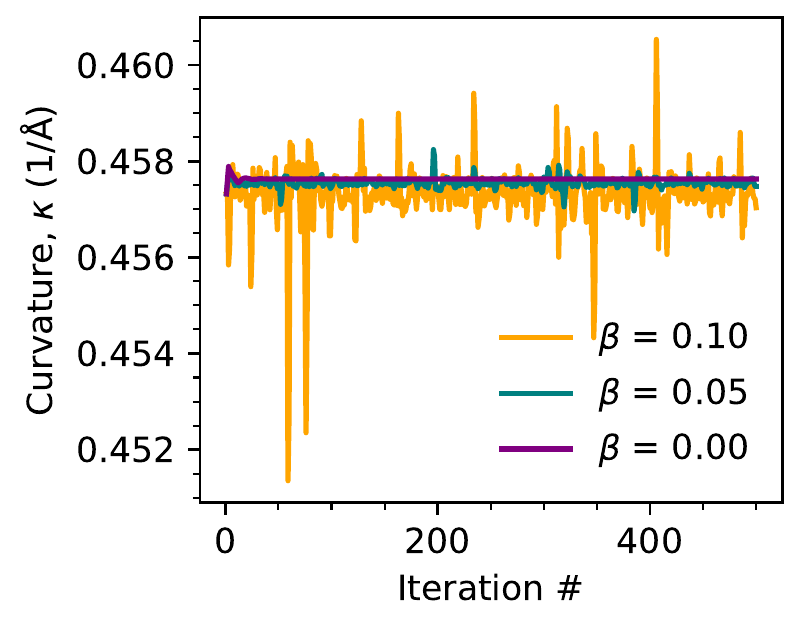}
\includegraphics[width=0.3\textwidth,clip]{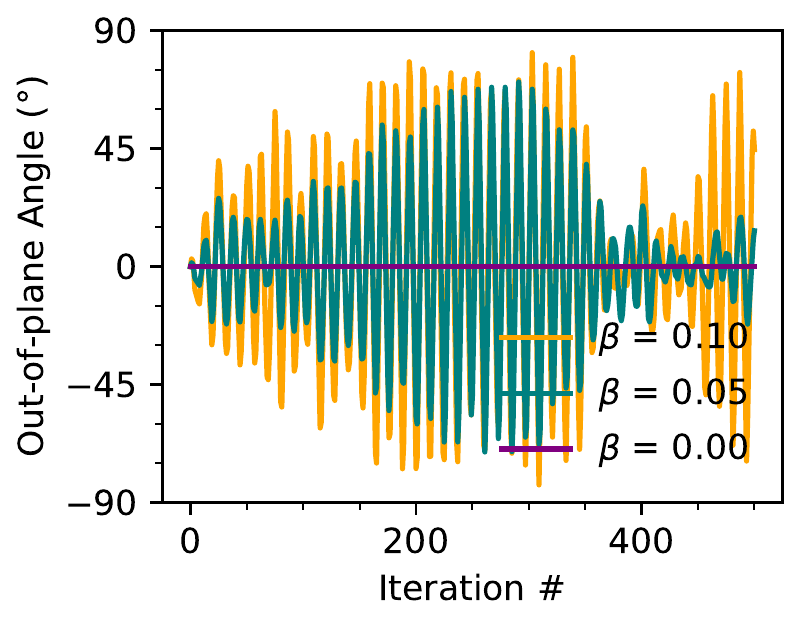}
\caption{\label{fig:two_atoms_performance} (top left) The potential energies relative to the target energy value with iteration number. (top middle) Histograms of potential energies relative to the target energy value. (top right) the step size. (bottom left) The distance between the two atoms of the dimer with the target shown in grey. (bottom middle) the computed curvature. (bottom right) The angle of the dimer relative to the rotational plane.}
\end{figure*}

Overall, there are four tunable parameters: maximum step $s_{max}$, angle limit $\phi$,  drift fraction $\beta$, and potentiostat step scale $\alpha$. The following heuristic for the potentiostat step scale is robust: $ \alpha = 1.1 + 0.6\,\beta$, 
which works because the drift vector is not informed by the curvature of the contour and it tends to move the system further from the energy target than is anticipated by extrapolation. We suspect the slight baseline overcorrection compensates for errors from the truncated Taylor expansion.
Increasing the potentiostat step scale proportionally with the drift lets this heuristic work well for many systems. 
A drift fraction of 0.1-0.2 generally works very quickly to break symmetries and orbits. Increasing the drift fraction above 0.2 begins weakening the potentiostatic accuracy of the position updates since the curvature extrapolation does not apply to the drift component. 
For small systems with a few atoms, the max step tends to be the most important to potentiostatic accuracy regardless of the angle limit. 
For systems with tens of atoms, the angle limit yields very good automatic tuning of step sizes and the max step can reliably be set quite high, e.g., $\sim$1-3 Å. 
Angle limits of $<30^\circ$ degrees are generally safe, but can be set much higher if the system has more uniform curvature and the drift fraction is low.  
Using a small angle limit and larger step max, tends to cause systems to take more small steps in corners and other highly curved energy boundaries, which may be a useful feature, for sampling configurational space. 

\begin{figure}
\centering
\includegraphics[width=0.220\textwidth,clip]{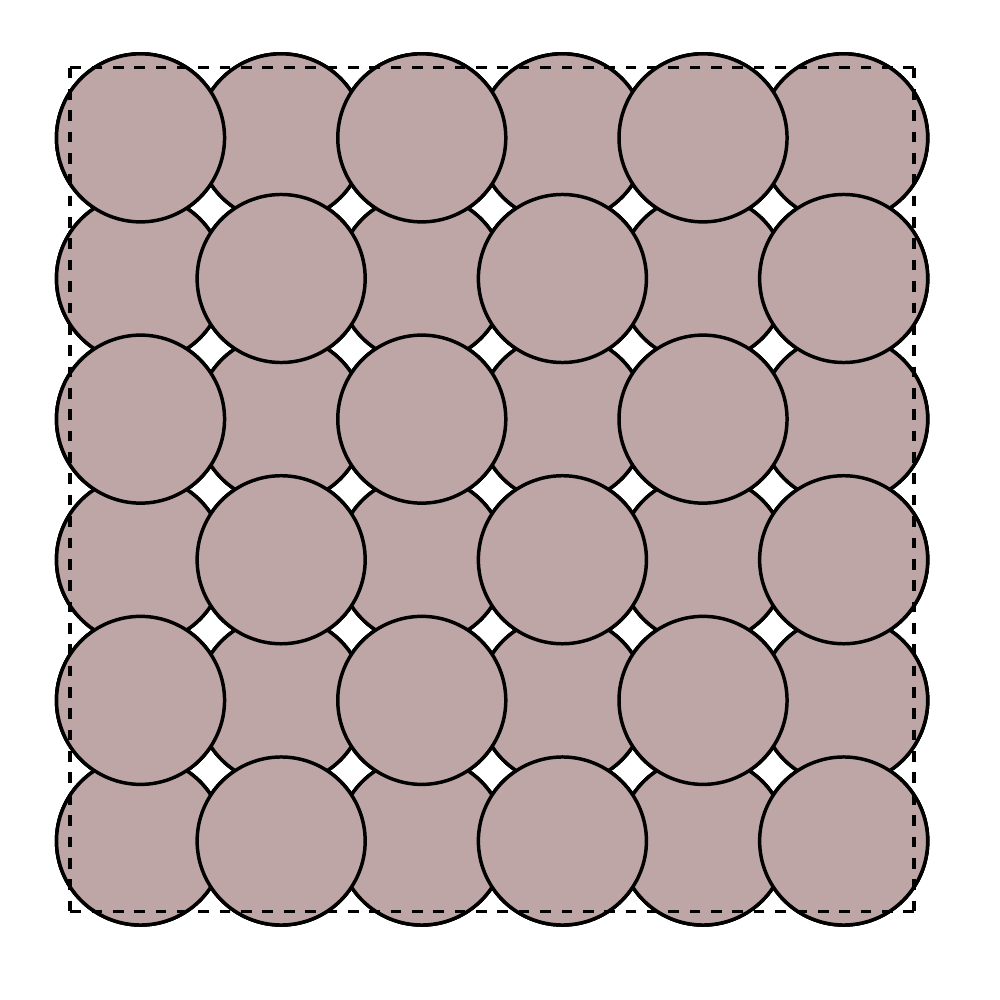}
\includegraphics[width=0.220\textwidth,clip]{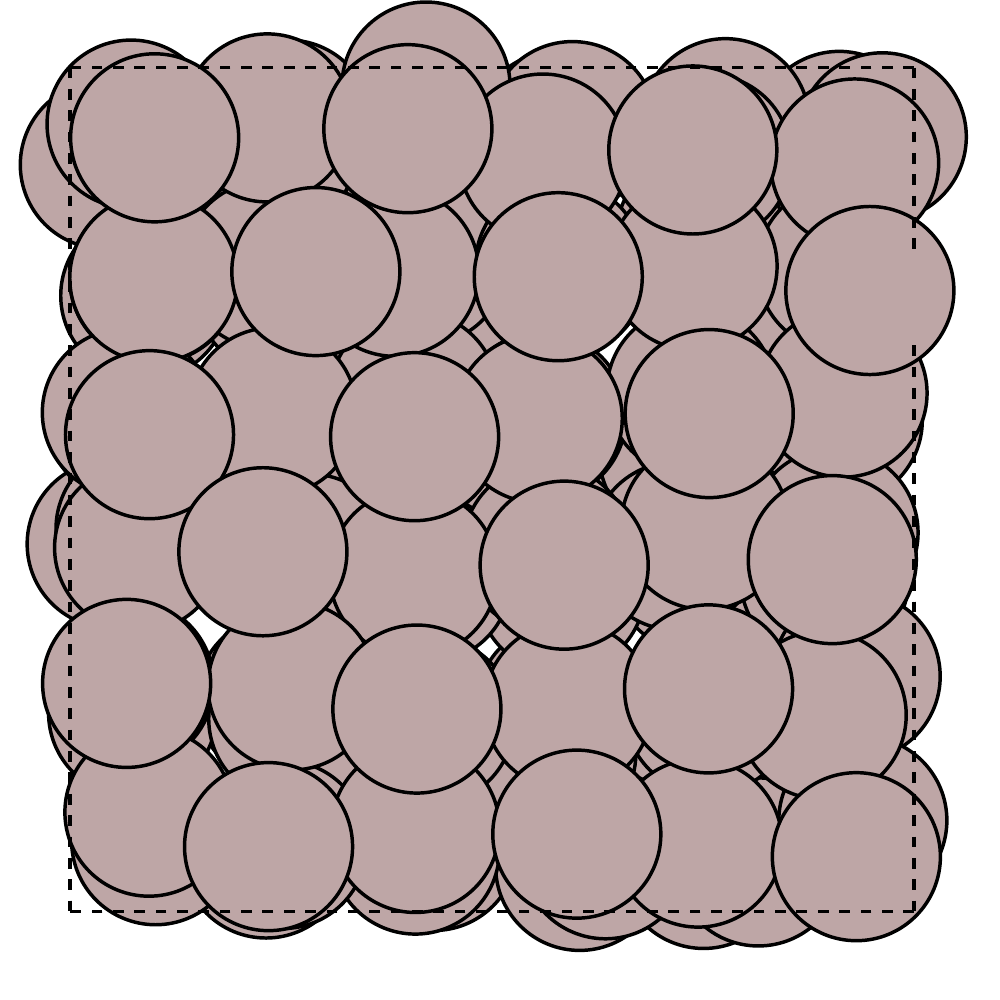}
\caption{\label{fig:bulk_crystal} The $3\times3\times3$ simulation cell of aluminum (left) in the ground state and (right) after 50 iterations of contour exploration at 5\,\% of its cohesive energy.
}
\end{figure}

\begin{figure*}
\centering
\includegraphics[width=0.30\textwidth,clip]{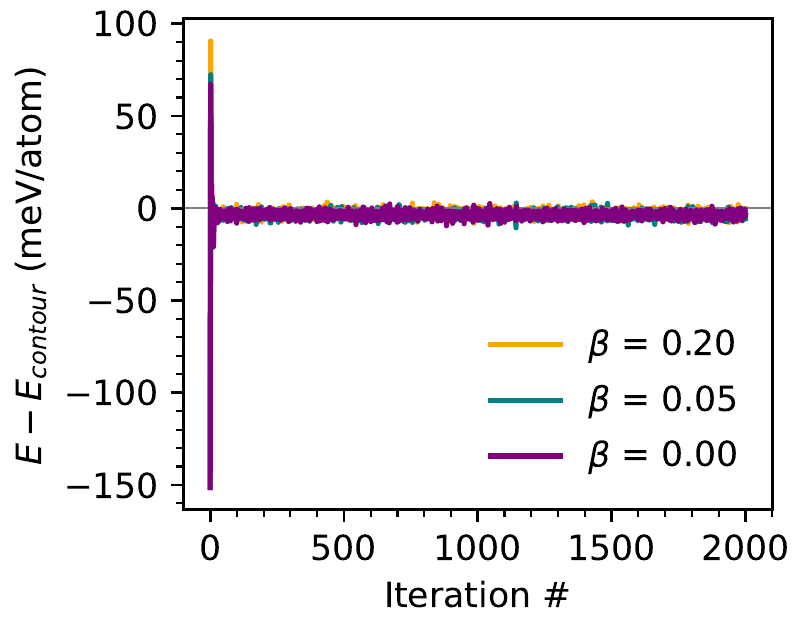}
\includegraphics[width=0.30\textwidth,clip]{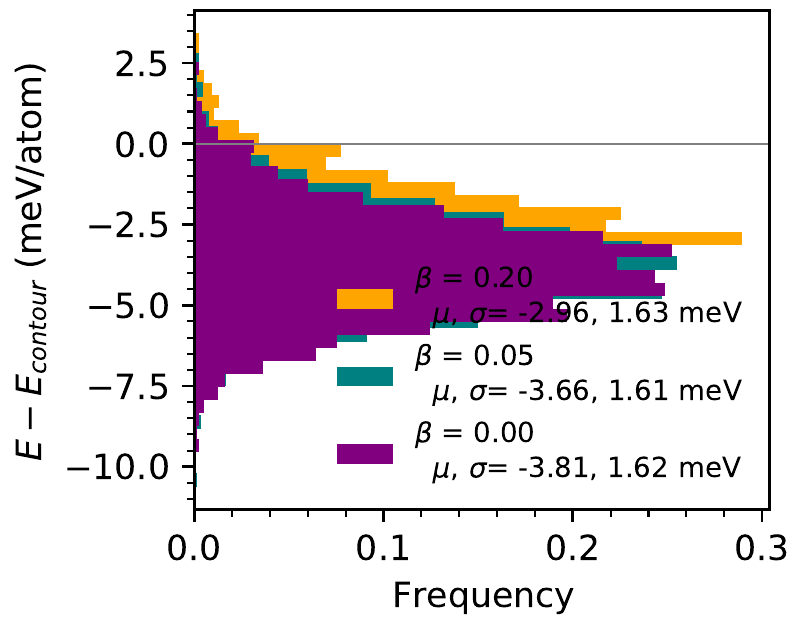}
\includegraphics[width=0.30\textwidth,clip]{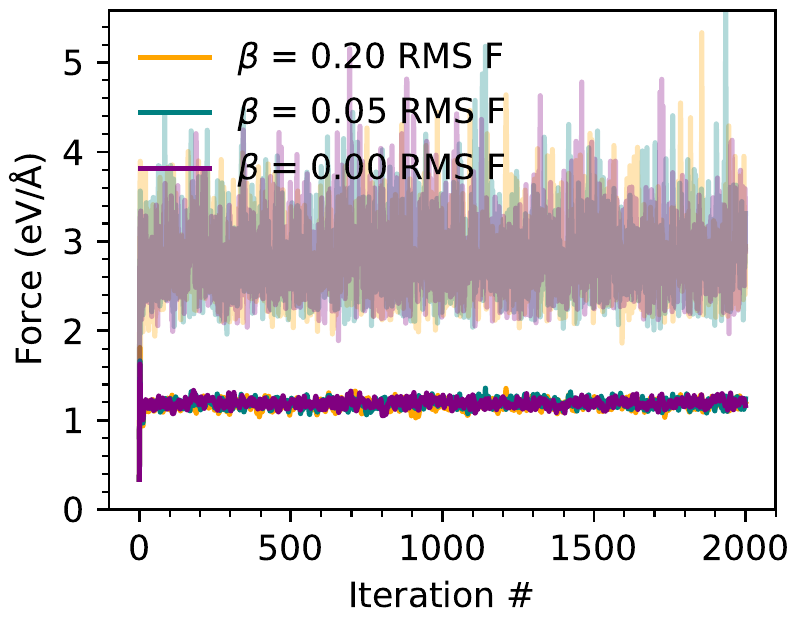}

\includegraphics[width=0.30\textwidth,clip]{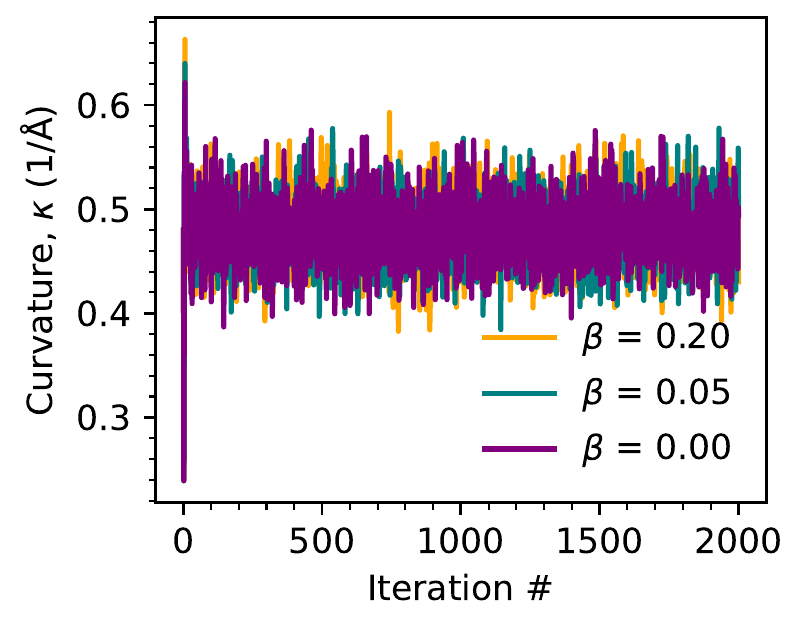}
\includegraphics[width=0.30\textwidth,clip]{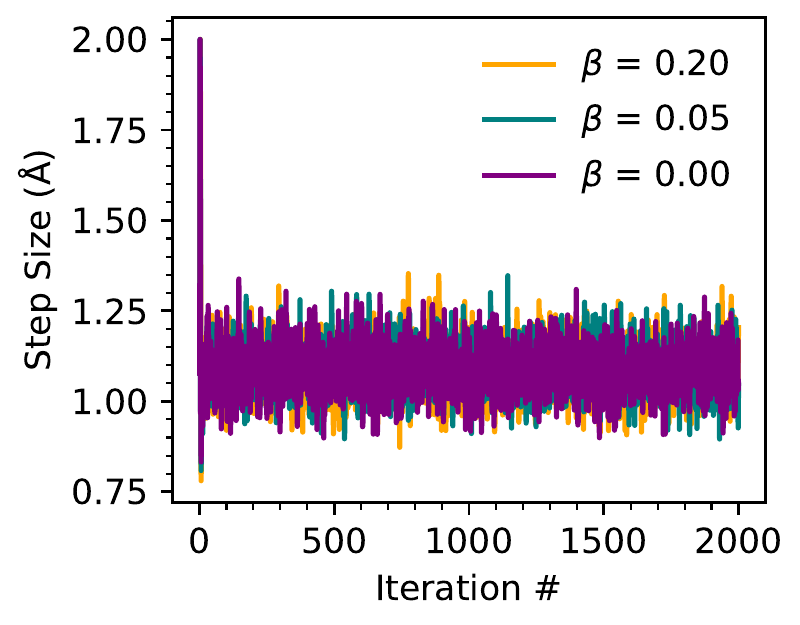}

\caption{\label{fig:bulk_crystal_stats} (top left) The potential energy relative to the energy target. (top center) the distributions of potential energy relative to the target energy. (top right) the RMS force and the maximum atomic forces (in pastel). (bottom left) the curvature of the contour. (bottom right) the step size used in the position update. 
}
\end{figure*}

\section{Demonstrations}
\subsection{Diatomic Molecule}
Since our method extrapolates to new points on a PEC using the local curvature, the best proof of concept is apply our method to a system with PECs of uniform curvature. 
In this case, an orbiting dimer of aluminum atoms which have equal and opposite initial velocity vectors to negate center of mass motion (\autoref{fig:two_atoms}). 
The potential energy of the atoms is modeled by the EMT interatomic potential provide with ASE.
The dimer has been stretched to 3.092 Å or 1.3$\times$ the equilibrium separation distance for a PEC 991.6 meV above equilibrium. 
The contour was explored using an angle limit of 30$^\circ$ and a max step of 2\,\AA. 
The first three iterations show the dimer rotating in the expected 30$^\circ$ (\autoref{fig:two_atoms}). 
The PEC was explored for 500 iterations using drift values of $\beta = 0, \, 0.05, \, 0.1$ to examine the affect of drift on potentiostatic accuracy. 
For all levels of drift, the potentiostat reaches a dynamic equilibrium after ~20 iterations and these iterations are not included in the statistical measures. 
The drift-free exploration desirably converges on a PEC only 2 meV/atom from the target PEC. 
Despite the large step size of 1.131 \AA, the position error of the PEC location is only 0.002 \AA. 
The PEC curvature is overestimated by a mere 0.06\%.
The inclusion of drift predictably results in potentiostatic errors, with errors growing rapidly with increasing drift fraction (\autoref{fig:two_atoms_performance} top left/center) in this low dimensionality test.
Moreover, the error distributions appear to have long tails. 
Yet without drift, the dimer may only rotate in-plane and it can be seen in \autoref{fig:two_atoms_performance} (bottom right) that a drift fraction of 0.1 quickly breaks the symmetry to allow the out-of plane PEC be explored. 
The errors can of course be tuned by reductions in an angle limit used in the contour exploration.

\subsection{Bulk Crystal}
Having shown that our methods work to follow PECs in a low-dimensional system, we now demonstrate its use for high dimensionality systems more representative of an \textit{ab initio} simulation of a bulk material. 
Using the same EMT interatomic potential, we apply our contour-exploration method to a $3\times3\times3$ conventional supercell of aluminum containing 108 atoms (\autoref{fig:bulk_crystal}). In contrast to the last example, we  start from the ground state and then use the potentiostat to raise the energy to a higher target value.   
This tasks leads to an important caveat of contour exploration: the method alone cannot be initiated from a ground state structure, since by definition there are no forces to travel perpendicular to when the system in its ground state. 
Additionally, the potentiostat has no means of increasing or decreasing energy if there are no forces to inform the potential energy gradient.

One remedy for simulating the potentiostatic kinematics from a ground state configuration is to apply small displacements in the  initialization.
In the aluminum supercell here, the small displacements are  normally-distributed with a standard deviation of 0.05\,\AA.
Now, we may proceed with the simulation with a target potential energy of 164.1 meV/atom, which is 5\,\% of the cohesive energy of aluminum in the interatomic potential model.
This potential energy target is high, but below the threshold where the individual atomic PECs merge into a single hyperplane. The merging of the individual atomic PECs permits free motion of atoms and the lattice is not preserved. 
The MD analogy would be keeping aluminum below its melting temperature. 
Starting from near ground state, the potential energy overshoots the target and equilibrates in a period of about 20 iterations as seen in \autoref{fig:bulk_crystal_stats}. 
Once equilibrated, the potentiostat systematically falls slightly below its potential energy target by 3-4 meV/atom with a narrow distribution of under 2 meV/atom regardless of the drift fraction used. 
The RMS forces stabilize just over 1 eV/Å while all forces are below 6 eV/Å. 
After the initial equilibration, the curvature oscillates near 0.5 Å$^{-1}$, which for the $30^\circ$ angle limit gives steps sizes near 1.1\,\AA.
This demonstration shows that our contour exploration method can explore high dimensional PECs with good potentiostatic accuracy. 

\begin{figure}[t]
\centering
\includegraphics[width=0.15\textwidth,clip]{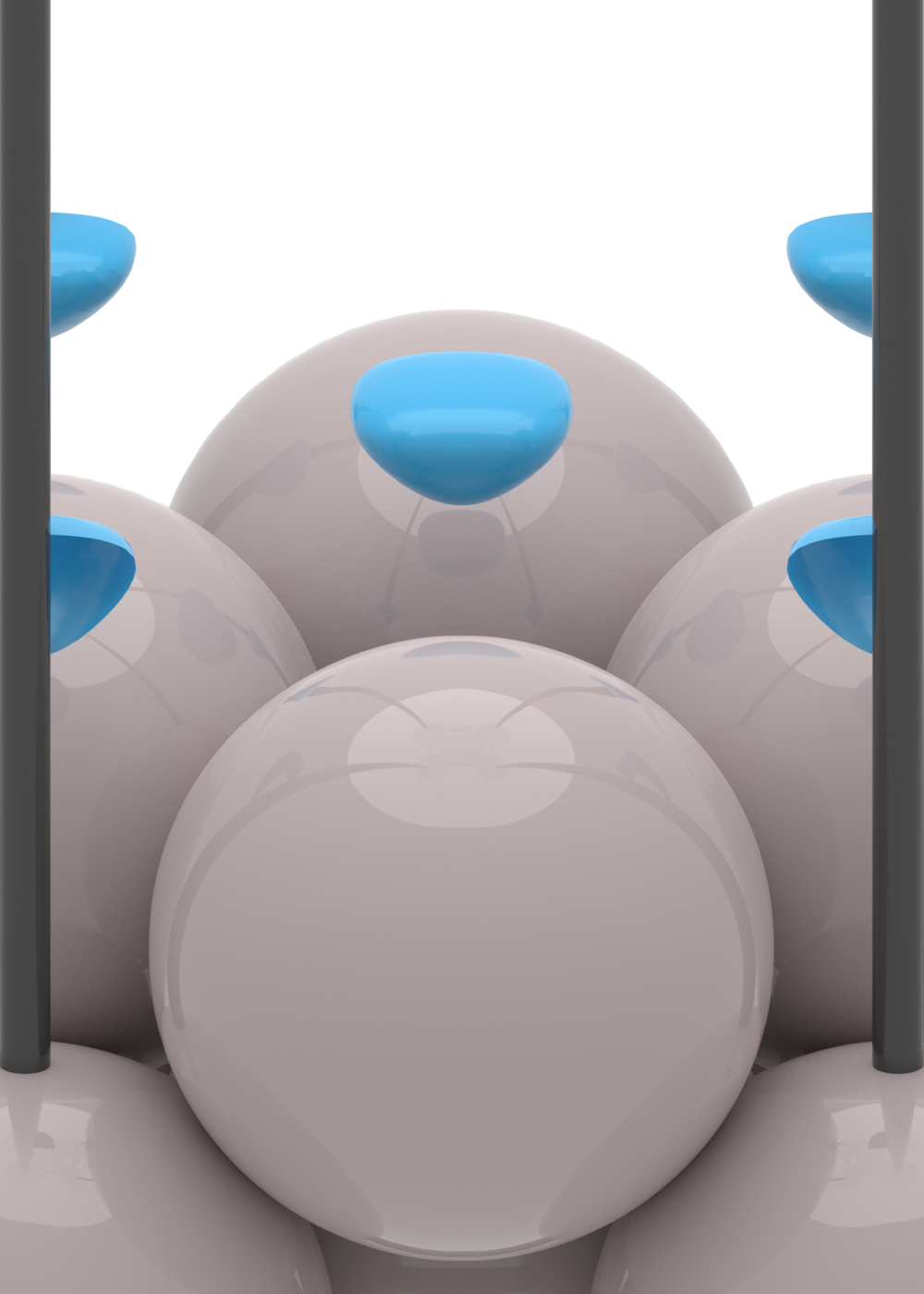}
\includegraphics[width=0.15\textwidth,clip]{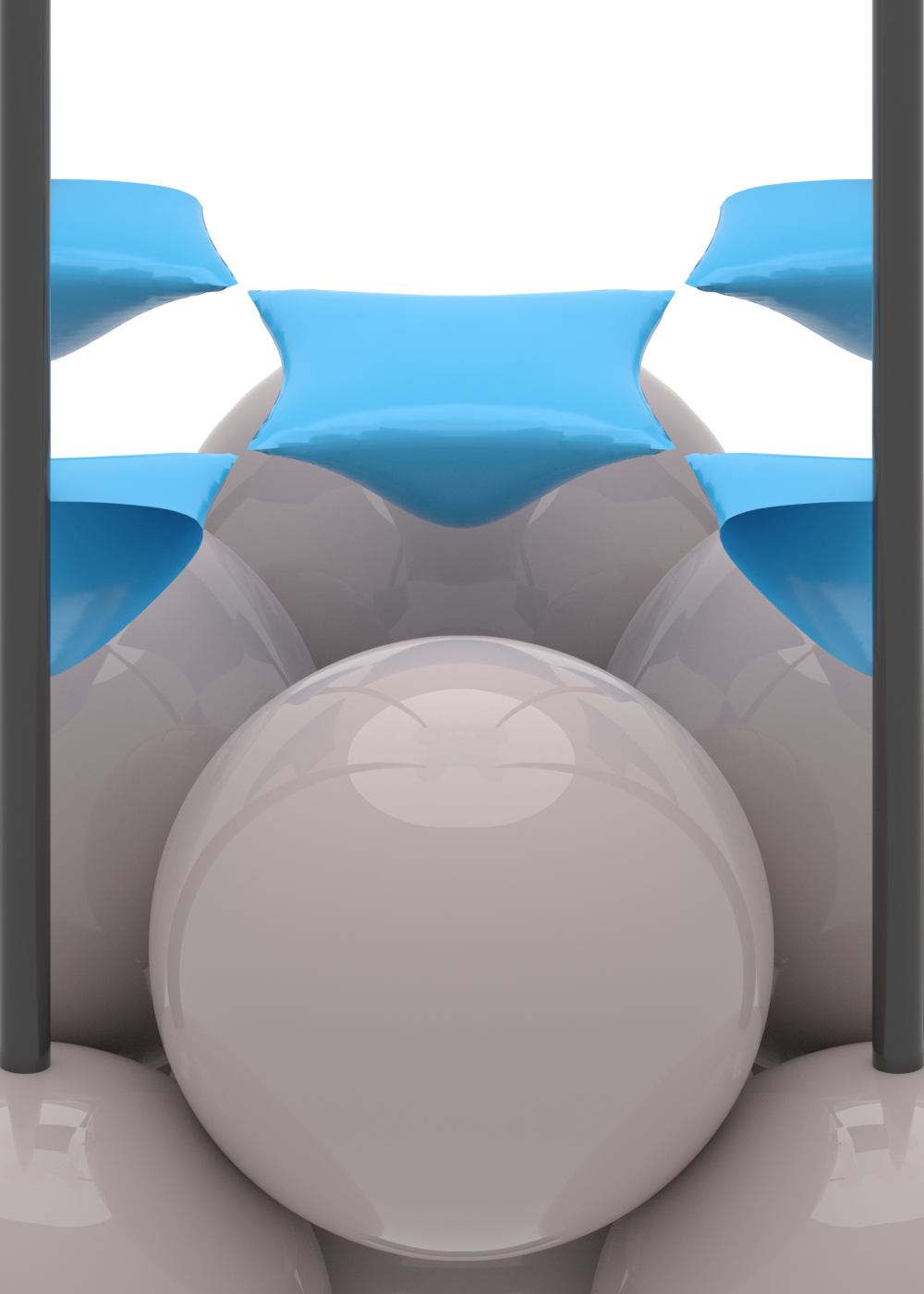}
\includegraphics[width=0.15\textwidth,clip]{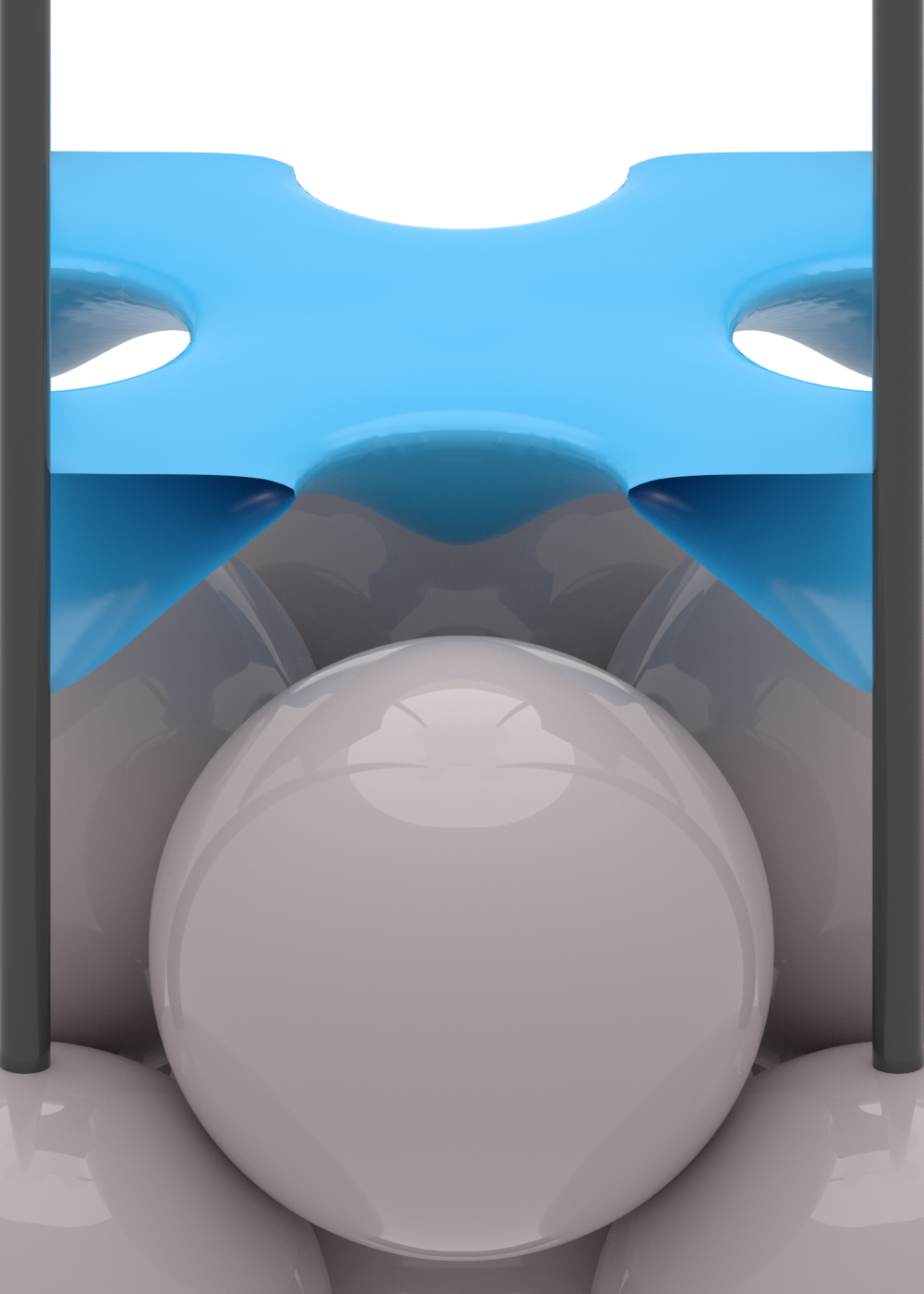}

\caption{\label{fig:isosurface} Potential energy contours for a copper atom on a rigid aluminum (001) surface at (left) 0.5X, (center) 1.0X, and (right) 1.5X the diffusion barrier energy of 319.65 meV are shown in blue. The potential energy contours edges are a result of the truncation to a single conventional cell. 
}
\end{figure}

\subsection{Atom on a Rigid Surface}
As previously mentioned, the angle limit of the position update directly scales the size of the update and, thus the potentiostatic accuracy. 
But the angle limit also modulates a bias in the sampled configuration space. 
With smaller angle limits for the step updates, contour exploration will preferentially sample areas of high curvature since it requires more steps to traverse them. 
To demonstrate curvature biased sampling, we examine PECs of a single copper atom traversing a fixed aluminum (100) surface. 
Importantly, using a single atom permits trivial visualization of the PECs for comparison (\autoref{fig:isosurface}).

\begin{figure}[t]
\centering
\includegraphics[width=0.185\textwidth,clip]{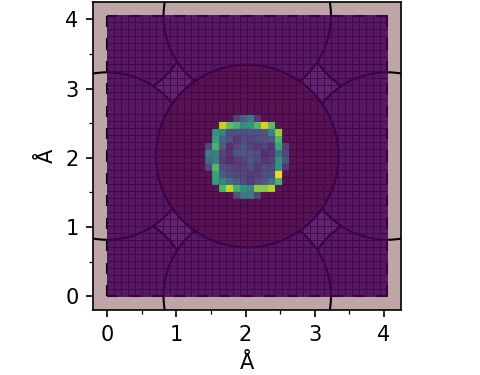}\hspace{-22pt}
\includegraphics[width=0.185\textwidth,clip]{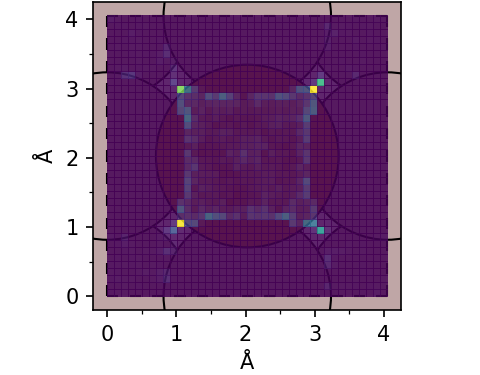}\hspace{-22pt}
\includegraphics[width=0.185\textwidth,clip]{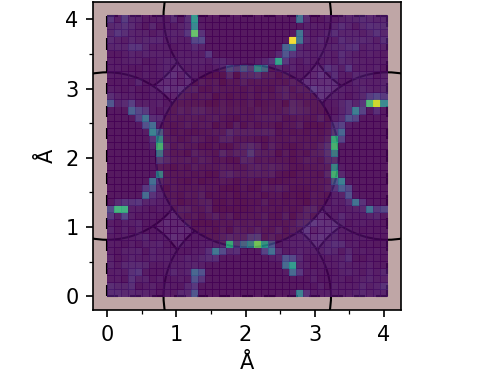}

\includegraphics[width=0.185\textwidth,clip]{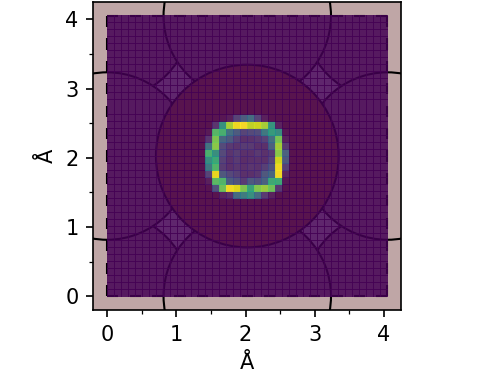}\hspace{-22pt}
\includegraphics[width=0.185\textwidth,clip]{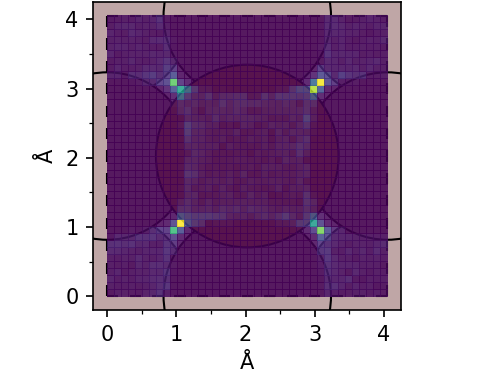}\hspace{-22pt}
\includegraphics[width=0.185\textwidth,clip]{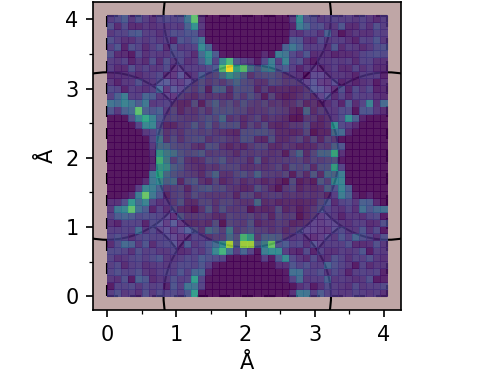}
\caption{\label{fig:surface_dist} 2D  histograms of the copper atom position during contour exploration on a frozen aluminum (001) surface for PECs at (left to right) 0.5X, 1.0X, and 1.5X the diffusion barrier energy  with angle limits of 5$^\circ$ (top) and 20$^\circ$ (bottom). 
}
\end{figure}

\begin{figure*}[t]
\centering
\includegraphics[width=0.3\textwidth,clip]{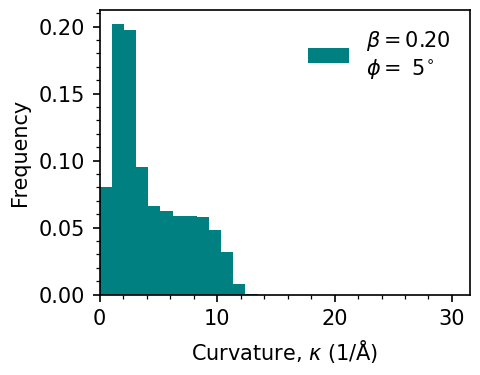}
\includegraphics[width=0.3\textwidth,clip]{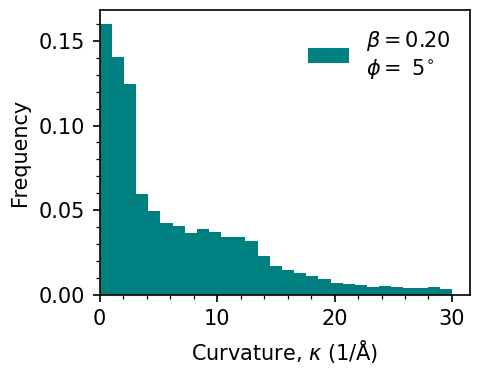}
\includegraphics[width=0.3\textwidth,clip]{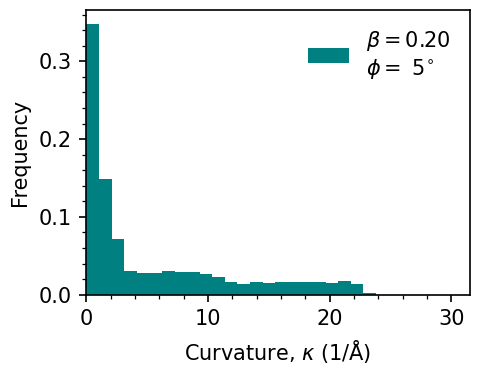}

\includegraphics[width=0.3\textwidth,clip]{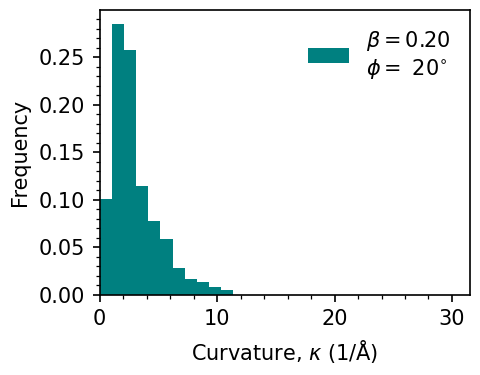}
\includegraphics[width=0.3\textwidth,clip]{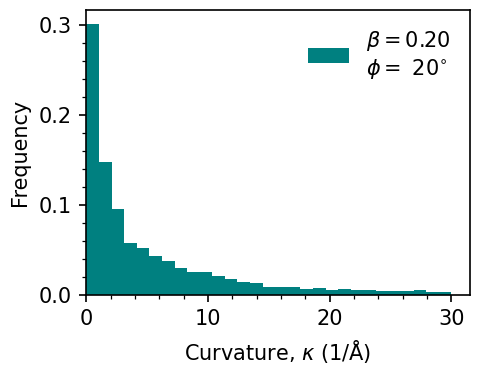}
\includegraphics[width=0.3\textwidth,clip]{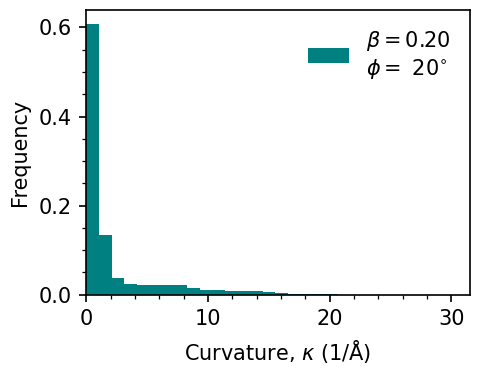}
\caption{\label{fig:curvature_hist} Potential energy contour curvature histograms for a copper atom on a frozen aluminum (001) surface for PECs at (from left to right) 0.5, 1.0, and 1.5x the diffusion barrier energy with angle limits of 5$^\circ$ (top) and 20$^\circ$ (bottom). 
}
\end{figure*}

Atomic motion and diffusion on surfaces, particularly the PES saddle points at the diffusion barriers are important for catalysis \cite{CompCatRev}.
As such, we perform contour exploration of the PECs at 0.5X, 1.0X, and 1.5X the diffusion barrier energy of 319.65 meV, corresponding to the barrier height for a displacement along the $\langle 100 \rangle$ directions.
As the potential energy increases, the initially separate PECs connect at the saddle point before forming a continuous surface (\autoref{fig:isosurface}).
In \autoref{fig:surface_dist}, these three PECs are explored with angle limits of 5$^\circ$ and 20$^\circ$ using 10,000 steps, a drift fraction of 0.2, and a maximum step of 0.2\,\AA. 
We find that a smaller angle limit leads a higher density of samples in areas of highest curvature on the PECs. 
This appears as well in the histograms in \autoref{fig:curvature_hist}, which show that a smaller angle limit results in more frequent sampling of higher curvature areas of the PEC at all energies.   
The PECs at the diffusion barrier energy have a curvature approaching infinity at the saddle point yielding practically very small steps, trapping the contour exploration near the saddle.   
Without \textit{a priori} knowledge of the saddle point energy, contour exploration is inefficient at finding saddle points because the very high curvature disappears quickly above and below the saddle point as shown in \autoref{fig:isosurface} for this example. 

The desirability of high sampling in areas of high curvature on the PEC is likely application dependent. 
For simply creating an accurate map of the potential energy contours, a more efficient representation can be achieved with curvature-biased sampling as a basic form of adaptive sampling \cite{arclength_sampling}.
For MLIP applications, a higher angle limit as seen in the bottom row of \autoref{fig:surface_dist} will likely prove more useful since the larger steps sizes will cover more configuration space and the distributions will be more uniform in that configuration space. 
These features are normally accessed by high temperature molecular dynamics simulations \cite{MLclassical, RosenbrockAlloys}, even when active learning is subsequently employed \cite{HafniaGap,MuellerLi}.

\section{Summary and Outlook}
We developed a potentiostatic kinematics algorithm  which is capable of following potential energy contours in systems with arbitrary dimensions. 
A simple potentiostat corrects for deviations from potential energy target.  
By using extrapolation based on the radius of curvature, the method can tune step size on-the-fly via a turning angle limit. 
The angle limit can be tuned to improve potentiostatic accuracy while increasing sampling in areas of high potential energy contour curvature. 
This approach may provide an alternative route  for creating robust training data sets for machine learned interatomic potentials, but there are likely uses outside this application.

Several additional improvements may enhance the versatility of potentiostatic kinematics, including:
\begin{enumerate}
\item The contour following extrapolation formula is based on the Taylor expansion of the Frenet-Serret formulas. 
The Frenet-Serret formulas are generalizable to $N$-dimensions \cite{FSblackholes}, which in conjunction with the trivial derivation of higher order Taylor expansions, gives two routes to explore for improving potentiostatic accuracy with higher order formulae.
For compatibility with the minimum two degrees of freedom, we only use the 3D Frenet-Serret extrapolation. For simplicity and to avoid issues with path dependency, we only use constant curvature extrapolation.
\item The accuracy the potentiostatic update may also be improved by moving beyond the linear extrapolation of the potential energy. For instance, the second derivative of the potential energy along the force direction could be derived from a Hessian estimating method that uses higher order optimization algorithms, \emph{e.g.},  the Broyden-Fletcher-Goldfarb-Shanno method.\cite{B, F, G, S}  
With the second derivative, the potentiostatic accuracy near harmonic minima would likely be improved.
\item For very large systems, widely separated atomic motions could be treated to independently contribute to the total potential energy. By spatially decomposing the system into multiple local curvatures to match, larger net steps sizes would likely be possible since the local position updates could be independently tuned to the local curvature. 
\item Lastly, in cases of systematic overshooting or undershooting of the energy target, an on-the-fly feedback mechanism could tune the scaling, $\alpha$, of the potentiostatic step size or shift the potential energy target. 
\end{enumerate}

\begin{acknowledgments}
This work relates to Department of Navy award N00014-20-1-2368 issued by the Office of Naval Research. The United States Government has a royalty-free license throughout the world in all copyrightable material contained herein.
\end{acknowledgments}

\bibliography{references}

\end{document}